\documentclass[10pt,towcolumn]{autart}

\date{\today}
\usepackage{subfigure}
\usepackage{comment}
\usepackage{empheq}
\usepackage{booktabs}
\usepackage{amsmath,amssymb}
\usepackage{graphicx}
\usepackage{multirow}
\usepackage{epstopdf}
\usepackage{float}
\usepackage{cite}
\usepackage{xcolor}
\usepackage{changes}
\allowdisplaybreaks[4]

\usepackage{appendix}
\usepackage{chngcntr}
\usepackage{color}

\def\DraftVersion{1} 

\begin{document}
\begin{frontmatter}
\title{Moving Horizon Estimation for ARMAX process with {\it t} -Distribution Noise\thanksref{footnoteinfo}}
\thanks[footnoteinfo]{The authors acknowledge support by the Singapore National Research Foundation (NRF) under its Campus for Research Excellence And Technological Enterprise (CREATE) programme, specifically the Cambridge Centre for Advanced Research and Education in Singapore (Cambridge CARES, http://www.cares.cam.ac.uk), project C4T. This paper was not presented at any IFAC
meeting. Corresponding author Dexiang Zhou. Tel. +65 8376 0221.}
\author[Dexiang Zhou]{Dexiang Zhou}\ead{ZHOU0180@e.ntu.edu.sg},    
\author[Dexiang Zhou]{Keck Voon Ling}\ead{EKVLING@ntu.edu.sg},               
\author[Weng Khuen Ho]{Weng Khuen Ho}\ead{wk.ho@nus.edu.sg},  
\author[Jan M. Maciejowski]{Jan M. Maciejowski}\ead{jmm@eng.cam.ac.uk}           \address[Dexiang Zhou]{School of Electrical $ \& $ Electronic Engineering, Nanyang Technological University, Singapore}
\address[Weng Khuen Ho]{Department of Electrical $\&$ Computer Engineering, National University of Singapore, Singapore }
\address[Jan M. Maciejowski]{Department of Engineering, University of Cambridge, Trumpington Street, Cambridge CB2 1PZ, UK}

\begin{abstract}
In this paper, instead of the usual Gaussian noise assumption, $t$-distribution noise is assumed. A Maximum Likelihood Estimator using the most recent N measurements is proposed for the Auto-Regressive-Moving-Average with eXogenous input (ARMAX) process with this assumption. The proposed estimator is robust to outliers because the `thick tail' of the t-distribution reduces the effect of large errors in the likelihood function. Instead of solving the resulting nonlinear estimator numerically, the Influence Function is used to formulate a computationally efficient recursive solution, which reduces to the traditional Moving Horizon Estimator when the noise is Gaussian.
The formula for the variance of the estimate is derived. This formula shows explicitly how the variance of the estimate is affected by the number of measurements and noise variance. The simulation results show that the proposed estimator has smaller variance and is more robust to outliers than the Moving Window Least-Squares Estimator. For the same accuracy, the proposed estimator is an order of magnitude faster than the particle filter.
\end{abstract}
\end{frontmatter}
\section{Introduction}
Gaussian noise is often assumed in state estimation. However, the Gaussian noise assumption is an approximation to reality. The occurrence of outliers, transient data in steady-state measurements, instrument failure, model nonlinearity, etc. can all induce non-Gaussian data \cite{wang2003framework}.

It is well-known that the $\it t$-distribution has the property of `\,thick tail'\, to better model the occurence of outliers \cite{kotz2004multivariate}. In addition, as a special case, the $\it t$-distribution reduces to the Gaussian distribution when its degree of freedom tends to infinity. Thus the $\it t$-distribution has  the flexibility to characterize noise with Gaussian or non-Gaussian statistical properties. In the literature the heavy tail property of the $\it t$-distribution has been used to improve the performance of estimators in \cite{meinhold1989robustification,agamennoni2011outlier,agamennoni2012approximate,roth2013student}. These papers are based on the Bayesian method and the heavy tail property of $t$-distribution was used to reduce the effect of  outliers. Our method differs from these estimators in that the probability density function (pdf) of the $\it t$-distribution is used directly in the likelihood function for a Maximum Likelihood Estimator (MLE).

 The sensitivity of an estimator when the underlying noise assumption (such as Gaussian noise) is violated has been extensively studied in the robust statistics literature \cite{huber1964robust,hampel2011robust,huber1981robust,maronna2006robust}. In particular, Huber \cite{huber1964robust} studied the effect of outliers by contaminating the underlying noise distribution with data from an arbitrary unknown distribution.   Hampel \cite{hampel2011robust} proposed the Influence Function (IF) approach to describe the effect of an infinitesimal contamination of the underlying noise distribution.
 If the IF of an estimator is bounded and/or decreasing, or increasing slowly for large magnitude of noise, then the estimator is robust to outliers.
 It can be shown that the IF of the least-squares estimator increases linearly with the magnitude of noise and is unbounded \cite{hampel2011robust}. This confirms the well-known fact that standard least-squares estimation is not robust to outliers. Other approaches for handling non-Gaussian noise include the particle filter \cite{arulampalam2002tutorial} which is based on the Monte Carlo method, but at the expense of a heavier computational load.

 IF has been mainly used as an analysis tool. The main contribution of this paper, Theorem \ref{theoremrecursiveMHET}, is to employ the IF for state estimation. We formulate a computationally efficient recursive algorithm that gives an approximate solution to the moving horizon MLE for Auto-Regressive-Moving-Average with eXogenous inputs (ARMAX) process with $t$-distribution noise. 


 ARMAX processes are popularly used in the statistical analysis of time series \cite{aastrom2011computer}. The Kalman filter for an ARMAX process is well-known, see Example 11.6 in \cite{aastrom2011computer}. In \cite{ho2014filtering} a MLE of an ARMAX process with generalized $t$-distribution noise was developed. However, in \cite{ho2014filtering} the IF was used to approximate the MLE by using all the measurements.
 This makes the estimator insensitive to plant parameter changes \cite{ling1999receding}.
 In this paper we derive the moving horizon version of the ARMAX filter. 
 We derive a computationally efficient recursive algorithm in Section 3. 
 We also analyze the statistical properties of the proposed estimator in Section 4. More specifically, the formula for the variance of the estimate is derived.

\section{Maximum Likelihood Estimation and its approximation for ARMAX process with $t-$distribution noise}
\label{SecIFexamplesimple}
Consider the following single-input single-output ARMAX process:
\begin{eqnarray}
\label{armaxtf}
A\left(q^{-1}\right)y_k=B\left(q^{-1}\right)u_k+C\left(q^{-1}\right)e_k,
\end{eqnarray}
where $k$ is the sampling instance, $u_k$ and $y_k$ are the input and output respectively. The polynomials are assumed to be co-prime and given as
\begin{eqnarray}
A\left(q^{-1}\right)&=&1+a_1q^{-1}+\cdots+a_nq^{-n}, \notag \\
B\left(q^{-1}\right)&=&b_1q^{-1}+b_2q^{-2}+\cdots+b_{n_B}q^{-n_B}, \notag \\
C\left(q^{-1}\right)&=&1+c_1q^{-1}+\cdots+c_nq^{-n}, \notag
\end{eqnarray} where $n_B\leqslant n$ and $q^{-1}$ is the backward shift operator, i.e., $q^{-1}y_k=y_{k-1}$. {\color{black}The zeros of the polynomials $A(q^{-1})$ and $C(q^{-1})$ are inside the unit disc. }
$e_k$ is an independent and identically distributed random variable associated with the zero mean $t$-distribution pdf \cite{kotz2004multivariate}
\begin{eqnarray}
   f(e_k) = \frac{ {\bf \Gamma}\left( \frac{\nu+1}{2} \right) }{\sqrt{\nu \pi} \sigma{\bf \Gamma}\left( \frac{\nu}{2}\right) } \left(1+\frac{e_k^2}{\sigma^2\nu}\right)^{-\frac{\nu+1}{2}}
   \triangleq t_{\nu}\left(0,\sigma\right)
   \label{ek}
\end{eqnarray}
  The degree of freedom and scale parameters are given by $\nu$ and $\sigma$ respectively, and ${\bf \Gamma}$ is the gamma function.

For convenience, we will derive the estimator in state-space form. The ARMAX process \eqref{armaxtf} in minimal state-space form is given as~\cite{aastrom2011computer}
\begin{subequations}
\label{armaxss}
\begin{eqnarray}
x_{k+1}&=&\Phi_a x_k+\Gamma u_k+\Omega e_k, \label{armaxssx}  \\
y_k&=&Hx_k+e_k, \label{armaxssy}
\end{eqnarray}
\end{subequations}
%

where
\begin{eqnarray}
\Phi_a&=&\left[\begin{matrix}
-a_1&1&0&0&\cdots&0&0 \\
-a_2&0&1&0&\cdots&0&0 \\
\vdots&\vdots&\vdots&\vdots&\ddots&\vdots&\vdots \\
-a_{n-1}&0&0&0&\cdots&0&1 \\
-a_n&0&0&0&\cdots&0&0
\end{matrix}\right]~\Gamma=\left[\begin{matrix}
b_1 \\
b_2\\
\vdots \\
b_{n_B} \\
0
\end{matrix}\right], \notag \\
\Omega&=&\left[\begin{matrix}
c_1-a_1 &
c_2-a_2 &
\cdots &
c_n-a_n
\end{matrix}\right]^{'}, H=\left[\begin{matrix}
1&0&\cdots&0
\end{matrix}\right]. \notag
\end{eqnarray}
\subsection{Maximum Likelihood Estimation}
Given measurements $y_1,~y_2,~\cdots,~y_{_{T}}$, we denote the MLE of $x_1$ for the ARMAX process \eqref{armaxss} with $t$-distribution noise as
\begin{eqnarray}
  \hat{x}_{_{\rm MLE}} = \arg~\!\min_{\hat x_1}
   J,  
   \label{MLEall}
\end{eqnarray}
where
\begin{subequations}
\begin{eqnarray}
J       &=& -\textstyle{\sum_{k=1}^{T}}\ln f(e_k),\label{J} \\
e_k&=&y_k- H\Phi^{k-1}\hat x_1-Hs_k,\label{MLE}  \\
\Phi&=&\Phi_a-\Omega H, \label{Phi} \\
s_k&=& \begin{cases}
{\bf 0}& {\rm if}~k = 1, \\
\textstyle{\sum_{i=1}^{k-1}}\Phi^{k-i-1}\left(\Gamma u_{i}+\Omega y_{i}\right)&{\rm if}~k\geqslant  2.
\end{cases} \label{xbar}
\end{eqnarray}
\end{subequations}
 Then the solution to \eqref{MLEall} can be found by solving the equation
\begin{eqnarray}
\textstyle{\sum_{k=1}^{T}}\psi_k(e_k) = 0, \label{psi}
\end{eqnarray}
where $\psi_k(e_k)\triangleq\frac{d}{d \hat x_1} \left(-\ln f\left(e_k\right)\right)=-\frac{\left(\nu+1\right)\left(H\Phi^{k-1}\right)^{'} e_k}{\sigma^2\nu+(e_k)^2}$.


\begin{defn}
\label{defIFhampel}
\textsc{Influence Function \cite{hampel2011robust}.}\\ The {\rm IF} of an estimator $G$ at  distribution $F$ is
\begin{eqnarray}
{\rm IF}\left(y;G,F\right)=\mathop{\lim}_{h\rightarrow 0}\frac{G\left((1-h)F+h\Delta_y\right)-G(F)}{h}, \notag
\end{eqnarray}
where $h\in [0,1]$ and $\Delta_y$ denotes the probability measure which puts mass 1 at the point $y$.
\end{defn}

The IF describes the effect of an infinitesimal contamination at the point $y$ on the estimate, standardized by the mass of the contamination. In this paper the noise is assumed to be $t$ distributed with pdf $f(e_k)$. Taking the expectation of \eqref{psi} gives
\begin{eqnarray}
   \textstyle{\sum_{k=1}^{T}} \textstyle{\int_{-\infty}^{\infty}} \psi_k\left(e_k\right)f\left(e_k \right)de_k = 0
\label{expectsimple2}.
\end{eqnarray}
Now we wish to obtain the IF of our estimator. As a first step, we replace $f(e_k)$ in \eqref{expectsimple2} with $(1-h)f\left(e_k \right)+h\Delta_{e_k}$
\begin{eqnarray}
   \textstyle{\sum_{k=1}^{T} \int_{-\infty}^{\infty}} \psi_k\left(e_k\right)\left[(1-h)f\left(e_k \right)+h\Delta_{e_k}\right]de_k = 0
\label{expectsimple}.
\end{eqnarray}
Equation \eqref{expectsimple} implicitly defines our estimator
as a function of $h$ and we denote it as $\hat x_1(h)$. In general, $\hat x_1(h)$ is a nonlinear state estimator. The Taylor series expansion of $\hat x_1(h)$ at $h=0$ is given by
\begin{eqnarray}
\hat x_1(h)= \hat x_1(0)+\textstyle{\left.{d \hat x_1(h) \over d h}\right|_{h=0}h}+\textstyle{\frac{1}{2} \left.\frac{d^2\hat x_1(h)}{dh^2}\right|_{h=0}h^2}+\cdots \notag \\ \label{Taylorseries}
\end{eqnarray}

Definition \ref{defIFhampel} suggests that the IF of our estimator can be obtained by differentiating equation (\ref{expectsimple}) wrt. $h$ and setting $h=0$
\begin{eqnarray}
   && {\rm IF}(e)=\left.{d \hat x_1(h) \over d h}\right|_{h=0} \notag \\ &=&-\left(\textstyle{\sum_{k=1}^{T}\int_{-\infty}^{\infty}  }
          \frac{d\psi_k\left(e\right)}{d \hat x_1}
              f\left(e\right)de\right)^{-1}
        \left.\textstyle{\sum_{k=1}^{T}} \psi_{k}\left(e_k\right)\right|_{\hat x_1=\bar x_1}.
        \label{IF}
\end{eqnarray}
where we denote $\hat{x}_1(0)$ as $\bar{x}_1$.

\if\DraftVersion 0
The derivation of \eqref{IF} is given in Appendix A in \cite{zhou2017moving}.
\else
The derivation of \eqref{IF} is given in Appendix \ref{APPIF}.
\fi


\subsection{{\rm IF} Approximation}
\begin{lem}
\label{propxIFt}

The {\rm IF} of MLE for ARMAX process \eqref{armaxss} with $t$-distribution noise \eqref{ek} is
\begin{eqnarray}
    {\rm IF}(e) &=&\left(\textstyle{\sum_{k=1}^{T}\left(H\Phi^{k-1}\right)^{'}H\Phi^{k-1}}\right. \notag \\
    &&\left.\times\textstyle{\int_{-\infty}^{\infty} }
          \frac{\left(\nu+1\right)\left(\sigma^2\nu-e^2\right)}
          {\left(\sigma^2\nu+e^2\right)^2}
              f\left(e\right)de\right)^{-1}\notag \\
              &&\times \left.\textstyle{\sum_{k=1}^{T}} \frac{\left(\nu+1\right)\left(H\Phi^{k-1}\right)^{'} e_k}{\sigma^2\nu+(e_k)^2}\right|_{\hat x_1=\bar x_1}. \label{IFt}
\end{eqnarray}
When the noise $e_k$ is Gaussian, 
\eqref{IFt} becomes
\begin{eqnarray}
 \mbox{\rm IF}(e) &=&-\bar x_1+\left(\textstyle{\sum_{k=1}^T}\left(H\Phi^{k-1}\right)^{'}H\Phi^{k-1}\right)^{-1}\notag \\
&& \times \textstyle{\sum_{k=1}^{T}}\left(H\Phi^{k-1}\right)^{'}\left(y_k-Hs_k\right). \label{IFgaussian}
\end{eqnarray}

\end{lem}
\begin{pf}
\if\DraftVersion 0
See Appendix B in \cite{zhou2017moving}.
\else
See Appendix \ref{APPpropxIFt}.
\fi
\end{pf}

Since \eqref{expectsimple} reduces to \eqref{psi} when $h=1$, $\hat x_1(1)$ is the solution that we are looking for, namely the solution to \eqref{psi}. We can get an approximate solution to this by retaining only the first two terms of \eqref{Taylorseries}. Using \eqref{IF},
\begin{eqnarray}
\label{xIF}
\hat x_{_{\rm MLE}} \approx \bar x_1+{\rm IF}(e)\triangleq x_{_{\rm IF}}.
\end{eqnarray}

%
\begin{rem}
When the noise is Gaussian, substituting \eqref{IFgaussian} into \eqref{xIF} gives  \begin{eqnarray}
 x_{_{\rm IF}} &=&\bar x_1+{\rm IF}(e)
=\left(\textstyle{\sum_{k=1}^T}\left(H\Phi^{k-1}\right)^{'}H\Phi^{k-1}\right)^{-1}\notag \\
&&\times\textstyle{\sum_{k=1}^{T}}\left(H\Phi^{k-1}\right)^{'}\left(y_k-Hs_k\right)
\label{xIFgaussian}
\end{eqnarray}
which is the least-squares estimator. It is re-assuring that our estimator reduces to the well-known result when the noise is Gaussian. Note that $ x_{_{\rm IF}}=\hat x_{_{\rm MLE}}$ in this case since under the Gaussian noise assumption, the MLE and least-squares estimator coincide.
\end{rem}

In the sequel we shall assume that $\bar x_1=0$, for simplicity. Then, since $\left.e_k\right|_{\hat x_1=\bar x_1}=y_k-H s_k$, \eqref{IFt} becomes
 \begin{eqnarray}
 {\rm IF}(e)&=&\left(\textstyle{\sum_{k=1}^{T}\left(H\Phi^{k-1}\right)^{'}H\Phi^{k-1}}\right.\notag \\
 &&\left.\times\textstyle{\int_{-\infty}^{\infty}}\frac{\left(\nu+1\right)\left(\sigma^2\nu-e^2\right)}{\left(\sigma^2\nu+e^2\right)^2}f\left(e\right)de\right)^{-1}\notag \\
 &&\times  \textstyle{\sum_{k=1}^{T}}\frac{(\nu+1)\left(H\Phi^{k-1}\right)^{'} \left(y_k-Hs_k\right)}{\sigma^2\nu+\left(y_k-Hs_k\right)^2}
 \label{IFxbar0}
 \end{eqnarray}
 and \eqref{xIF} becomes
  \begin{eqnarray}
 \label{xIFxbar0}
 x_{_{\rm IF}}={\rm IF}(e).
 \end{eqnarray}

 \begin{rem}
We consider only cases when $C\left(q^{-1}\right)$ is asymptotically stable. In this case, $\bar x_1=0$ is a simplifying assumption since error due to a wrong initial state estimate will reduce to zero after an initial transient.
%
\end{rem}

\section{The Proposed Estimator}
\label{secMLE}

Instead of using all the measurements, we adopt the moving horizon principle and use the most recent $N$ measurements $\left\lbrace y_k\right\rbrace_{T-N+1}^T$.
\if\DraftVersion 1
Therefore, the IF of MHE-TD can be obtained by replacing $k=1$ with $k=T-N+1$ in \eqref{IFxbar0}
\begin{eqnarray}
{\rm IF}(e)&=& \left(\textstyle{\sum_{\substack{k=T-\\N+1}}^{T}}\left(H\Phi^{k-1}\right)^{'}H\Phi^{k-1}\right.\notag\\
&&\left.\times\int_{-\infty}^\infty{(\nu+1)\left(\sigma^2\nu-e^2\right) \over \left(\sigma^2\nu+e^2\right)^2}f(e)de\right)^{-1}\notag\\
&&\times\textstyle{\sum_{k=T-N+1}^{T}}{(\nu+1)\left(H\Phi^{k-1}\right)^{'}\left(y_k-Hs_k\right) \over \sigma^2\nu+\left(y_k-Hs_k\right)^2}.
\label{IFarmaxMHE}
\end{eqnarray}
\fi
We denote the proposed estimator as Moving Horizon Estimation of ARMAX process with $t$-distribution noise (MHE-TD). By recursively applying the ARMAX model, we can obtain an expression for the estimate of the current state $x_{_{T}}$, and this is given in the next theorem.



\begin{thm}
\label{propMHETD}
\textsc{IF approximation of Batch MHE-TD}~~~ \rm

Given the ARMAX process \eqref{armaxtf} where $C(q^{-1})$ is asymptotically stable and noise $e_k$ with $\it t$-distribution pdf of \eqref{ek}, the IF approximation of MHE-TD is
\begin{subequations}
\label{xyhat}
\begin{eqnarray}
\hat{x}_{_{T}}&=&\textcolor{black}{s}_{_{T}}+\left(\textstyle{\sum_{i=1}^{N}}\left(H\Phi^{i-N}\right)'H\Phi^{i-N}\right)^{-1}\varPsi \mathcal W_{T,N},\label{xhat}\\
 \hat{y}_{_{T}}&=&H\hat{x}_{_{T}}, \label{yhat}
\end{eqnarray}
\end{subequations}
where $s_{_T}$ is given by \eqref{xbar} and
\begin{subequations}
\begin{eqnarray}
\varPsi&=&\left[\begin{matrix}
\left(H\Phi^{1-N}\right)^{'}&\left(H\Phi^{2-N}\right)^{'}&\cdots&H^{'}
\end{matrix}\right], \label{Psi}\\
\mathcal W_{T,N}&=&\left[\begin{matrix}
w_{_{T-N+1}}&w_{_{T-N+2}}&\cdots&w_{_{T}}
\end{matrix}\right]^{'}, \label{nonw}\\
w_{_{T-N+i}}&=&\left(\textstyle{\int\limits_{-\infty}^{\infty}}\frac{\left(\nu+1\right)\left(\sigma^2\nu-e^2\right)}{\left(\sigma^2\nu+e^2\right)^2}f(e)de\right)^{-1} \notag \\
&&\times \frac{\left(\nu+1\right)\left(y_{_{T-N+i}}-Hs_{_{T-N+i}}\right)}{\sigma^2\nu+\left(y_{_{T-N+i}}-Hs_{_{T-N+i}}\right)^2},~i=1,2,\cdots,N. \notag \\
\label{w}
\end{eqnarray}
\end{subequations}

\end{thm}
\begin{pf}
\if\DraftVersion 0
See Appendix C in \cite{zhou2017moving}.
\else
See Appendix \ref{APPpropMHETD}.
\fi
\end{pf}


In many cases the observations are obtained sequentially.
It is desirable to derive 
a recursive
solution for the IF approximation of MHE-TD as follows.
\begin{thm}
\label{theoremrecursiveMHET}
\textsc{IF Approximation of Recursive MHE-TD}~~~~The {\rm IF} approximation of MHE-TD satisfies the following recursive algorithm
\begin{subequations}
\label{MovingstrategyIF}
\begin{eqnarray}
\hat{x}_{_{T+1}}&=&\Phi \hat{x}_{_{T}}+\Gamma u_{_{T}}+\Omega y_{_{T}}\notag \\
&&+L_N\left(z_{_{T+1}}-H\left(\Phi\hat x_{_{T}}+\Gamma u_{_{T}}+\Omega y_{_{T}}\right)\right)\notag\\
&&-\tilde{L}_N\left(z_{_{T-N+1}}-H\Phi^{-N+1}\hat{x}_{_{T}}+H\Phi^{-N+1}\right.\notag \\
&&\left.\times\left(\varXi_uU_{T-1,N-1}+\varXi_yY_{T-1,N-1}\right)\right), \label{Movingstrategy1}  \\
\hat{y}_{_{T+1}}&=&H\hat{x}_{_{T+1}}, \label{Movingstrategy4}
\end{eqnarray}
\end{subequations}
where
\begin{subequations}
\begin{eqnarray}
 z_{k}&=&w_{k}+Hs_k, \label{zkp}  \\
 w_{k}
 &=&\left(\textstyle{\int\limits_{-\infty}^{\infty}}\frac{\left(\sigma^2\nu-e^2\right)}{\left(\sigma^2\nu+e^2\right)^2}f(e)de\right)^{-1}\frac{y_{k}-Hs_{k}}{\sigma^2\nu+\left(y_{k}-Hs_{k}\right)^2}, \notag \\ && k=T-N+1~{\rm or}~T+1, \label{wkp1} \\
L_N&=&\Phi^{N-1}P_N\left(H\Phi^{N-1}\right)^{'}, \label{LN} \\
\tilde{L}_N&=&\Phi^{N-1}P_N\left(H\Phi^{-1}\right)^{'}, \label{LNtilde} \\
P_N&=&\left(\textstyle{\sum_{i=1}^{N}}\left(H\Phi^{i-1}\right)^{'}H\Phi^{i-1}\right)^{-1},  \notag \\
s_{_{T+1}}&=&\Phi s_{_{T}}+\Gamma u_{_{T}}+\Omega y_{_{T}},  \notag \\
U_{T-1,N-1}&=&\left[\begin{matrix}
u_{_{T-N+1}}&u_{_{T-N+2}}&\cdots&u_{_{T-1}}
\end{matrix}\right]^{'}, \notag \\
Y_{T-1,N-1}&=&\left[\begin{matrix}
y_{_{T-N+1}}&y_{_{T-N+2}}&\cdots&y_{_{T-1}}
\end{matrix}\right]^{'}, \notag \\
\varXi_u&=&\left[\begin{matrix}
\Phi^{N-2}\Gamma&\Phi^{N-3}\Gamma&\cdots&\Gamma
\end{matrix}\right], \notag \\
\varXi_{y}&=&\left[\begin{matrix}
\Phi^{N-2}\Omega&\Phi^{N-3}\Omega&\cdots&\Omega
\end{matrix}\right]. \notag
\end{eqnarray}
\end{subequations}
\end{thm}
\begin{pf}
\if\DraftVersion 0
See Appendix D in \cite{zhou2017moving}.
\else
See Appendix \ref{APPtheoremrecursiveMHET}.
\fi
\end{pf}

\begin{rem}
\label{remIFlsr}
When the noise $e_k$ in ARMAX process \eqref{armaxtf} is Gaussian { \rm (i.e., $ \nu=\infty $)}, $w_k$ in \eqref{wkp1} reduces to $y_k-Hs_k$. Then $z_k$ reduces to $y_k$. Thus the recursive MHE-TD reduces to the following recursive solution of the standard Moving Window Least-Squares Estimator (MWLSE).
\end{rem}
\begin{subequations}
\label{Moving Windowleastsquaresestimation}
\begin{eqnarray}
\hat{x}_{_{T+1}}&=&\Phi \hat{x}_{_{T}}+\Gamma u_{_{T}}+\Omega y_{_{T}} \notag\\
&&+L_N\left(y_{_{T+1}}-H\left(\Phi\hat x_{_{T}}+\Gamma u_{_{T}}+\Omega y_{_{T}}\right)\right)\notag \\
&&-\tilde{L}_N\left(y_{_{T-N+1}}-H\Phi^{-N+1}\hat{x}_{_{T}}\right. \notag \\
&&\left.+H\Phi^{-N+1}\left(\varXi_uU_{T-1,N-1}+\varXi_yY_{T-1,N-1}\right)\right), \label{Moving Windowleastsquaresestimation1} \\
\hat{y}_{_{T+1}}&=&H\hat{x}_{_{T+1}}.\label{Moving Windowleastsquaresestimation2}
\end{eqnarray}
\end{subequations}
 Comparing \eqref{Movingstrategy1} and \eqref{Moving Windowleastsquaresestimation1} the recursive MHE-TD has the same structure as the standard MWLSE except for a nonlinear transformation, 
  i.e., \eqref{wkp1}. When $\Omega=0,~\Gamma=0$, Equation \eqref{Moving Windowleastsquaresestimation1} reduces to {\rm (10)} of \cite{ling1999receding}.
 \if\DraftVersion 1
The MHE uses only a finite number of past measurement samples; the oldest measurement is discarded as a new sample becomes available. The ARMAX filter \cite{ho2014filtering} uses the so-called growing memory strategy where the past measurement is not discarded as new measurement becomes available. Therefore, the ARMAX filter can be obtained if the last term of \eqref{Movingstrategy1} is discarded and $L_N$ and $P_N$ become
\begin{eqnarray}
L_T&=&\Phi^{T-1}P_T\left(H\Phi^{T-1}\right)^{'}, \notag \\
P_T&=&\left(\textstyle{\sum_{k=1}^{T}}\left(H\Phi^{k-1}\right)^{'}H\Phi^{k-1}\right)^{-1}. \notag
\end{eqnarray}
\fi

\section{Variance of the Proposed Estimator}
\if\DraftVersion 0
  Let the actual noise have pdf $g(e)$ which is different from $f(e)$ used in the design of MHE-TD. It is well-known that the mean and variance of an estimator can be analysed by its IF \cite{hampel2011robust, mises1947asymptotic, fernholz2001multivariate}. The statistical properties of the IF approximation of MHE-TD are derived by using IF as follows.
\else
 Let the actual noise have pdf $g(e)$ which is different from $f(e)$ used in the design of MHE-TD. It is well-known that the mean and variance of an estimator can be analysed by its IF \cite{hampel2011robust, mises1947asymptotic, fernholz2001multivariate}. The statistical properties of the IF approximation of MHE-TD are derived by using IF as follows.
\fi

\begin{thm}
\label{theorempropertyMHET}
\textsc{Variance of IF Approximation of MHE-TD }~~~ Consider the estimator \eqref{xyhat}. Assume that the data are generated from the ARMAX process \eqref{armaxtf} where the $t$-distribution pdf of $e_k$ is $g(e)$. 
The variance of the estimate is given by
\begin{eqnarray}
&&Var\left(\hat x_{_{T}}\right)\notag\\
&=&\textstyle{\frac{\rho_1}{\rho_4^2}}M_N+2{\frac{\rho_2}{\rho_4}M_N}\textstyle{\sum_{i=1}^{N-1}}\left(H\Phi^{i-N}\right)^{'}\left({\Phi_a}^{N-i-1}\Omega\right)^{'}\notag \\
&&+{\rho_3}\textstyle{\sum_{k=1}^{T-1}}{\Phi_a}^{T-k-1}\Omega\left({\Phi_a}^{T-k-1}\Omega\right)^{'}, \label{varxhat}
\end{eqnarray}
where
\begin{eqnarray}
\rho_1&=&\textstyle{\int_{-\infty}^{\infty}}\left(\frac{(\nu+1)e}{\sigma^2\nu + e^2}\right)^2g(e)de,~\rho_2=\textstyle{\int_{-\infty}^{\infty}}\frac{(\nu+1)e^2}{\sigma^2\nu + e^2}g(e)de, \notag \\
\rho_3&=&\textstyle{\int_{-\infty}^{\infty}}e^2g(e)de,~\rho_4=\textstyle{\int_{-\infty}^{\infty}}{(\nu+1)\left(\sigma^2\nu-e^2\right) \over (\sigma^2\nu+e^2)^2}g(e)de,\notag  \\
M_N&=&\left(\textstyle{\sum_{i=1}^{N}}\left(H\Phi^{i-N}\right)^{'}H\Phi^{i-N}\right)^{-1}. \notag
\end{eqnarray}
With $\hat y_{_{T}}=H\hat x_{_{T}}$,
\begin{eqnarray}
Var\left(\hat y_{_{T}}\right)&=&\textstyle{\frac{\rho_1}{\rho_4^2}}HM_NH^{'}+2\textcolor{black}{\frac{\rho_2}{\rho_4}HM_N}\textstyle{\sum_{i=1}^{N-1}}\left(H\Phi^{i-N}\right)^{'}\notag\\
&&\times H\textcolor{black}{\Phi_a}^{N-i-1}\Omega \notag \\
&&+\textcolor{black}{\rho_3}\textstyle{\sum_{k=1}^{T-1}}\left(H\textcolor{black}{\Phi_a}^{T-k-1}\Omega\right)^{2}. \label{varyhat}
\end{eqnarray}
\end{thm}
\begin{pf}
\if\DraftVersion 0
See Appendix E in \cite{zhou2017moving}.
\else
See Appendix \ref{APPtheorempropertyMHET}.
\fi
\end{pf}
The third term of \eqref{varyhat} which depends on $T$ is due to the dynamic of the ARAMX model.
For a fixed $T$, when $N$ increases $M_N$ decreases. Thus the variance decreases with increasing $N$. This is also shown in Table \ref{tabletN31}.

\if\DraftVersion 1

\begin{cor}
	\label{corollarypropertyMHEgaussian}
	The expectation of Moving Window Least-Squares Estimation is given by
	\begin{eqnarray}
	&&E\left(\hat x_{_{T}}\right)=\sum_{k=1}^{T-1}\left(\Phi+\Omega H\right)^{T-k-1}\Gamma u_k \notag
	\end{eqnarray}
	and its variance is
	\begin{eqnarray}
	&&Var\left(\hat x_{_{T}}\right)\notag \\
	&=&\rho_3\left(M_N+2\sum_{i=1}^{N-1}M_N\left(H\Phi^{i-N}\right)^{'}\left(\left(\Phi+\Omega H\right)^{N-i-1}\Omega\right)^{'}\right.\notag\\
	&&\left.+\sum_{k=1}^{T-1}\left(\Phi+\Omega H\right)^{T-k-1}\Omega\left(\left(\Phi+\Omega H\right)^{T-k-1}\Omega\right)^{'}\right), \notag \\
	\label{varxhatgaussian}
	\end{eqnarray}
	where
	\begin{eqnarray}
	\rho_3=\int_{-\infty}^{\infty}e^2g(e)d(e).
	\end{eqnarray}
	
	With $\hat y_{_{T}}=H\hat x_{_{T}}$,
	\begin{eqnarray}
	&&E\left(\hat y_{_{T}}\right)=H\sum_{k=1}^{T-1}\left(\Phi+\Omega H\right)^{T-k-1}\Gamma u_k \notag \\
	&&Var\left(\hat y_{_{T}}\right)\notag \\
	&=&\rho_3\left(HM_NH^{'}+2\sum_{i=1}^{N-1}HM_N\left(H\Phi^{i-N}\right)^{'}\right.\notag \\
	&&\left.\times H\left(\Phi+\Omega H\right)^{N-i-1}\Omega+\sum_{k=1}^{T-1}\left(H\left(\Phi+\Omega H\right)^{T-k-1}\Omega\right)^2\right).\notag \\
	 \label{varyhatgaussian}
	\end{eqnarray}
\end{cor}
\begin{pf}
	By setting $\nu=\infty$ in \eqref{varxhat}.
\end{pf}

\fi

The assumption that $g(e)$ is the same at each sampling instant $k$ is commonly made. In the following we extend to the case where $g(e)$ could be different at different $k$ denoted by $g_k(e)$. This is useful for the analysis of outliers (see Example 2).

\begin{thm}
	\label{corollaryMHEoutlier}
	\textsc{Property of IF approximation of MHE-TD  with outlier}~~~ {\replaced{ \textcolor{black}{Consider the estimator \eqref{xyhat}. Assume that the data are generated from the ARMAX process \eqref{armaxtf} where $x_1=0$ and} }{} the noise distribution contains an outlier at $k_1$, i.e.,
		\begin{eqnarray}
		g_k(e) = \begin{cases}
		\Delta_{e_{k_1}} & {\rm if}~k = k_1, \\
		g(e) &{\rm if}~k\neq k_1.
		\end{cases} \label{gkpdf}
		\end{eqnarray}
		then the expectation of the estimator \eqref{xyhat} is
		\begin{eqnarray}
		E\left(\hat y_{_{T}}\right)=\left\{
		\begin{aligned}
		&HE\left(s_{_{T}}\right)+\textstyle{\frac{1}{\rho_4}}HM_N\frac{\left(H\Phi^{k_1-T}\right)^{'}(\nu+1)e_{k_1}}{\sigma^2\nu+e_{k_1}^2} ~\notag \\
		&~~~~~~~~~~~~~~~~~~~{\rm if}~~k_1 \leq T \leq k_1+N-1,   \\
		&HE\left(s_{_{T}}\right)~~~~~~{\rm if}~T<k_1~{\rm or}~T>k_1+N-1,
		\end{aligned}
		\right.  \\
		\label{meanyhatoutlier}
		\end{eqnarray}
		where
		\begin{eqnarray}
		\label{meanxbaroutlier}
		E\left(s_{_{T}}\right)=\left\{
		\begin{aligned}
		&\textstyle{\sum_{k=1}^{T-1}}{\Phi_a}^{T-k-1}\Gamma u_k~~~~~~~~~{\rm if}~T\leq k_1, \\
		&\textstyle{\sum_{k=1}^{T-1}}{\Phi_a}^{T-k-1}\Gamma u_k +{\Phi_a}^{T-k_1-1}\Omega e_{k_1} \notag \\
		&~~~~~~~~~~~~~~~~~~~~~~~~~~~~~~~~~~~~~{\rm if}~~T> k_1. \notag
		\end{aligned}
		\right. \notag \\
		\end{eqnarray}}
\end{thm}
\begin{pf}
\if\DraftVersion 0
See Appendix F in \cite{zhou2017moving}.
\else
See Appendix \ref{APPcorollaryMHEoutlier}.
\fi
	
\end{pf}
\begin{rem}
	The expectation of Moving Window Least-Squares Estimation with outlier $e_{k_1}$ can be obtained by setting $\nu=\infty$ in \eqref{meanyhatoutlier}
	\begin{eqnarray}
	E\left(\hat y_{_{T}}\right)=\left\{
	\begin{aligned}
	&HE\left(s_{_{T}}\right)+HM_N\left(H\Phi^{k_1-T}\right)^{'}e_{k_1} \notag \\
	&~~~~~~~~~~~~~~~~~~~~~~~~~{\rm if}~k_1 \leq T \leq k_1+N-1,   \\
	&HE\left(s_{_{T}}\right) ~~~~~~~~~~~~{\rm if}~T<k_1~{\rm or}~T>k_1+N-1,
	\end{aligned}
	\right. \notag \\
\label{meanyhatoutliergaussian}
	\end{eqnarray}
	where $E\left(s_{_{T}}\right)$ is given in \eqref{meanxbaroutlier}.
\end{rem}

\section{Examples}
\label{secSimulations}
The models and parameters used in the examples are summarized in Table \ref{Table:Table1}. The statistical results in Example 1,2 and 3 are obtained from 1000 Monte Carlo simulation runs.

 {\footnotesize
\begin{table}
\small
  \caption{Models and Parameters Used in the Examples}\label{Table:Table1}
  \centering
  \begin{tabular}{|c|c|c|c|}
    \hline
    Example & 1 &  2  & 3  \\ \hline
    $N$ & 6,9,12 & \multicolumn{2}{|c|}{3}  \\ \hline
    $e_k$ & $t_3(0,0.5)$ & \multicolumn{2}{|c|}{$t_3(0,1)$} \\ \hline
    $A(q^{-1})$ & $(1+0.855q^{-1})^5$ & \multicolumn{2}{|c|}{$1-0.9q^{-1}$}   \\
    $B(q^{-1})$ & $0.1q^{-1}$ & \multicolumn{2}{|c|}{$q^{-1}$}   \\
    $C(q^{-1})$ & $(1+0.8q^{-1})^5$ & \multicolumn{2}{|c|}{$1-0.85q^{-1}$} \\

    \hline
  \end{tabular}

\end{table}
}

 {\color{black} \subsection{Example 1:Estimation  Variance}

\label{secSimulationvariance2}


 Table \ref{tabletN31} shows the variances from 1000 simulation runs. The proposed estimator (Equation \eqref{MovingstrategyIF}), ARMAX filter \cite{ho2014filtering}, Kalman filter and MWLSE (Equation \eqref{Moving Windowleastsquaresestimation}) are compared for different horizon length, $N$, and different instance, $k$.
 \if\DraftVersion 0
 \else
  An example used to illustrate the calculation can be found in Appendix \ref{APPExample}.
 \fi

 Firstly, it can be seen that for the proposed estimator the theoretical and simulation variance matched approximately. Secondly, when $N$ increases, the variance decreases and therefore $N$ can be used as a tuning parameter for estimator performance. Thirdly, the variance of the proposed estimator is less than that of the MWLSE. Fourthly, for small $k$ ($=6,9,12$) our proposed estimator gives smaller variances than the Kalman Filter because of more accurate noise modeling. However, for large $k$ ($=40,60$), the estimators with growing memory (ARMAX filter and Kalman filter) give smaller estimate variance than the estimators with fixed memory (the proposed estimator and MWLSE). Finally, since the proposed estimator is the moving horizon version of the ARMAX filter, the results for both are the same when $N=k$. Likewise, the MWLSE is the moving horizon version of the Kalman filter, the results for both are also the same for $N=k$.

%
%
%
%

 {\footnotesize
 \begin{table}[!htb]
 	\small
 	\caption{Variance of $\hat y_{_{k}}$ in Example 1 }
 	\label{tabletN31}
 	\tabcolsep=0.05cm

 	\begin{tabular}{|c|c|c|c|c|c|c|c|}
 		\hline
 		
 		\multicolumn{3}{|c|}{$ k $}&6&9&12&40&60 \\
 \hline
 		\multirow{7}{1.5cm}{{Proposed estimator} Eq. \eqref{MovingstrategyIF}}&\multirow{4}{1.5cm}{Theoretical results from  \eqref{varyhat}}&N&&&&&\\
 		\cline{3-3}
 		&&$6$&0.6124&0.7055&0.7675&0.8617&0.8620\\
 		&&$9$&-&0.6879&0.7499&0.8441&0.8444\\
 		&&$12$&-&-&0.7237&0.8180&0.8183\\
 		\cline{2-8}
 		&\multirow{3}{1.5cm}{Simulation results}&$6$&0.5946&0.6834 &0.7416 &0.8486 &0.8729 \\
 		&&$9$&-&0.6723&0.7543 &0.8338 &0.8546 \\
 		&&$12$&-&-&0.7065 &0.7881 &0.8343 \\
 \hline		
 		
 		\multirow{3}{1.5cm}{MWLSE Eq. \eqref{Moving Windowleastsquaresestimation}}
 		&\multirow{3}{1.5cm}{Simulation results}
 		&$6$&0.9773&1.0705&1.1320&1.2270&1.2270\\
 		&&$9$&-&1.0360&1.0980&1.1920&1.1920\\
 		&&$12$&-&-&1.0460&1.1400&1.1400\\ \hline
        &&&&&&& \\[-3mm]
 		\multirow{1}{1.5cm}{ARMAX filter\cite{ho2014filtering}} &\multirow{1}{1.5cm}{Simulation results}&-&0.5946&0.6723 &0.7065 &0.4883&0.4889\\ [-3mm]
 &&&&&&& \\
 \hline		
 &&&&&&& \\[-3mm]
 		\multirow{1}{1.5cm}{Kalman filter }&\multirow{1}{1.5cm}{Simulation results}&-&0.9773&1.0360&1.0460&0.4881&0.4889\\ [-3mm]
&&&&&&& \\
 		\hline
 		
 	\end{tabular}
 \end{table}
}

  Figure \ref{figestimationerror} shows that the estimation error of the proposed estimator is smaller than that of MWLSE.

 \begin{figure}[!htb]
 	\centering
 	\includegraphics[width=8cm,height=6cm]{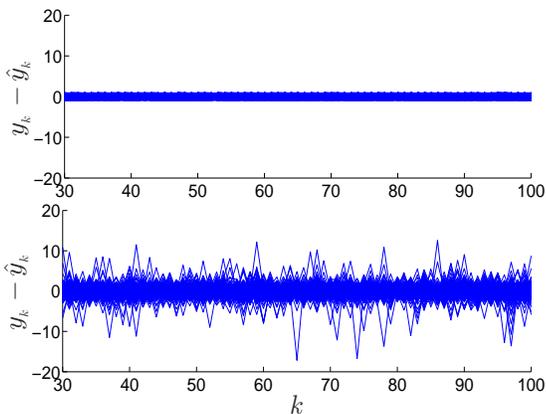}
 \vspace{-3mm}
 	\caption{Example 1: Estimation error: (Top) the proposed estimator with $N=6$; (Bottom) MWLSE with $N=6$.}
 	\label{figestimationerror}
 \end{figure}
 }

\subsection{Example 2: Performance with outlier}

 An outlier of $-10$ is introduced at $k=30$.

\begin{figure}[!htb]
\centering
\includegraphics[width=8cm,height=6cm]{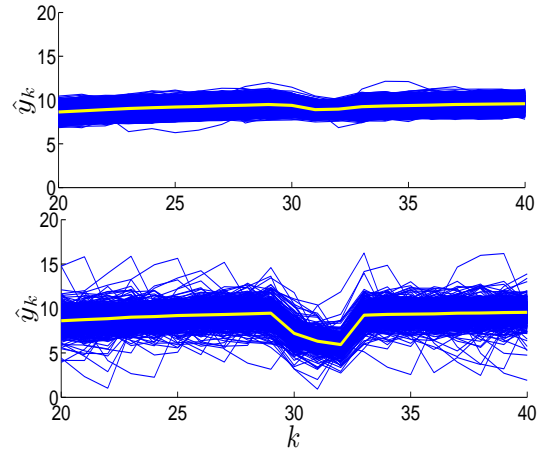}
\vspace{-3mm}
\caption{Example 2: (Top) $\hat{y}_k$ of the proposed MHE-TD estimator; (Bottom) $\hat{y}_k$ of MWLSE.}
\label{figIFlso}
\end{figure}

Figure~\ref{figIFlso} shows 1000 simulation runs of the proposed estimator (Equation \eqref{MovingstrategyIF}) and MWLSE (Equation \eqref{Moving Windowleastsquaresestimation}).  The yellow curves are the mean values of the 1000 runs. The proposed estimator is less affected by the outlier than the MWLSE. These yellow curves in the top and bottom figures can be calculated from \eqref{meanyhatoutlier} and \eqref{meanyhatoutliergaussian} respectively.
\subsection{Example 3: Comparison with particle filter}
The proposed estimator is compared with a particle filter {(bootstrap implementation)}\cite{arulampalam2002tutorial}.
Taking the numerical solution of MLE \eqref{MLEall} as the ground truth, we compared the estimators using the index
     \begin{equation}
     \notag
     I_k = \textstyle{\frac{1}{1000} \sum_{j=1}^{1000} \left\|\hat x_k^{(MLE,j)}-\hat x_k^{(j)}\right\|_2^2}
     \end{equation} where, for the $j$-th run, $\hat x_k^{(MLE,j)}$ is the numerical solution of the MLE \eqref{MLEall} and $\hat x_k^{(j)}$ is the estimate of $x_k$.
The indexes, at $k=50$, are $0.0082$, $0.0142$ and $0.0016$ for the proposed estimator, and particle filters with 100 and 1000 particles respectively.
The computational times per run using MATLAB R2015a with i7-5500U processor @2.4GHz, 16 GB RAM, are of the order of $0.1$ millisecond, one millisecond and ten milliseconds respectively. For about the same performances ($0.0082$ and $0.0142$) as given by index $I$, the computational time of the proposed estimator is an order of magnitude faster than the particle filter.

\section{Conclusion}\label{sec:con}

 The IF is employed to give an approximate solution to the moving horizon MLE for ARMAX process with $t$-distribution noise. The approximate solution can be formulated as a recursive MHE-TD algorithm which makes it suitable for on-line and real-time implementation at high sampling rates. We also used the IF to derive a formula for the estimates.  
  The examples show that the proposed estimator gives an estimate with a {smaller variance} than the MWLSE. It is also less affected by outliers than the MWLSE. 
  For about the same performance index, the computational time of the proposed estimator is an order of magnitude faster than the particle filter.


 \bibliographystyle{plain}
 \bibliography{autoreference}

\if\DraftVersion 1

\begin{appendices}
\appendix

\section{Appendix: Derivation of \eqref{IF}}
\label{APPIF}
Differentiating Equation (\ref{expectsimple}) wrt. $h$ gives
\begin{eqnarray}
&&\textstyle{\sum_{k=1}^{T}\int_{-\infty}^{\infty}}\frac{d\psi_k\left(e_k\right)}{dh}\left[(1-h)f\left(e_k\right)+h\Delta_{e_k}\right]de_k \notag \\
&&+\textstyle{\sum_{k=1}^{T}\int_{-\infty}^{\infty}}\psi_k\left(e_k\right)\left[-f\left(e_k\right)+\Delta_{e_k}\right]de_k=0. \label{gradientsimple1}
\end{eqnarray}
Substituting \eqref{expectsimple2} and
\begin{eqnarray}
\frac{d\psi_k\left(e_k\right)}{dh}=\frac{d\psi_k\left(e_k\right)}{d\hat x_1(h)}\frac{d \hat x_1(h)}{dh} \notag
\end{eqnarray}
into \eqref{gradientsimple1}  gives
\begin{eqnarray}
&&\textstyle{\sum_{k=1}^{T}\int_{-\infty}^{\infty}}\frac{d\psi_k\left(e_k\right)}{d\hat x_1(h)}\left[(1-h)f\left(e_k\right)+h\Delta_{e_k}\right]de_k\frac{d \hat x_1(h)}{dh}\notag \\
&&+\textstyle{\sum_{k=1}^{T}\int_{-\infty}^{\infty}}\psi_k\left(e_k\right)\Delta_{e_k}de_k=0. \label{gradientsimple5}
\end{eqnarray}
Setting $h=0$ in \eqref{gradientsimple5} gives
\begin{eqnarray}
&&\textstyle{\sum_{k=1}^{T}\int_{-\infty}^{\infty}}\frac{d\psi_k\left(e_k\right)}{d\hat x_1(h)}f\left(e_k\right)de_k\left.\frac{d \hat x_1(h)}{dh}\right|_{h=0}\notag \\
&&+\textstyle{\sum_{k=1}^{T}\int_{-\infty}^{\infty}}\psi_k\left(e_k\right)\Delta_{e_k}de_k=0. \notag
\end{eqnarray}
We denote $\hat x_1(0)$ as $\bar x_1$.

Thus we have
\begin{eqnarray}
\left.\frac{d \hat x_1(h)}{dh}\right|_{h=0}&=&-\left(\textstyle{\sum_{k=1}^{T}\int_{-\infty}^{\infty}}\frac{d\psi_k\left(e_k\right)}{d\hat x_1}f\left(e_k\right)de_k\right)^{-1}\notag \\
&&\times\left.\textstyle{\sum_{k=1}^{T}}\psi_k\left(e_k\right)\right|_{\hat x_1=\bar x_1}.
\label{IF2}
\end{eqnarray}
Replacing the integration variable $e_k$ by $e$ in \eqref{IF2} gives
\begin{eqnarray}
\left.\frac{d \hat x_1(h)}{dh}\right|_{h=0}&=&-\left(\textstyle{\sum_{k=1}^{T}\int_{-\infty}^{\infty}}\frac{d\psi_k\left(e\right)}{d\hat x_1}f\left(e\right)de\right)^{-1}\notag \\
&&\times \left.\textstyle{\sum_{k=1}^{T}}\psi_k\left(e_k\right)\right|_{\hat x_1=\bar x_1}.
\notag
\end{eqnarray}
which is \eqref{IF}.
\section{Appendix: Proof of Lemma \ref{propxIFt}}
\label{APPpropxIFt}
Substituting \eqref{ek} into \eqref{J}
\begin{eqnarray}
&&-{\rm ln}f\left(e_k\right)\notag \\
&=&-\ln\left(\frac{{\bf \Gamma}\left(\frac{\nu+1}{2}\right)}{\sqrt{\nu \pi}\sigma{\bf \Gamma}\left(\frac{\nu}{2}\right)}\right.\notag \\
&&\left.\times\left(1+\frac{\left(y_k-H\Phi^{k-1} \hat{x}_1-H\textcolor{black}{s}_k\right)^2}{\sigma^2\nu}\right)^{-\frac{\nu+1}{2}}\right) \notag \\
&=&-{\rm ln}\left(\frac{{\bf \Gamma}\left(\frac{\nu+1}{2}\right)}{\sqrt{\nu \pi}\sigma{\bf \Gamma}\left(\frac{\nu}{2}\right)}\right)\notag \\
&&+\frac{\nu+1}{2}{\rm ln}\left(1+\frac{\left(y_k-H\Phi^{k-1} \hat{x}_1-H\textcolor{black}{s}_k\right)^2}{\sigma^2\nu}\right). \notag
\end{eqnarray}

Differentiating $-{\rm ln}f\left(e_k\right)$ with respect to $\hat{x}_1$ gives
\begin{eqnarray}
&&\psi_k\left(e_k\right)=\frac{d}{d  \hat{x}_1} \left(-{\rm ln}f\left(e_k\right)\right)\notag \\
&=&\frac{\nu+1}{2}\frac{d}{d  \hat{x}_1}\left({\rm ln}\left(1+\frac{\left(y_k-H\Phi^{k-1} \hat{x}_1-H\textcolor{black}{s}_k\right)^2}{\sigma^2\nu}\right)\right)\notag \\
&=&-\frac{(\nu+1)\left(H\Phi^{k-1}\right)^{'}\left(y_k-H\Phi^{k-1} \hat{x}_1-H\textcolor{black}{s}_k\right)}{\sigma^2\nu+\left(y_k-H\Phi^{k-1} \hat{x}_1-H\textcolor{black}{s}_k\right)^2}
\notag \\
&=&-\frac{(\nu+1)\left(H\Phi^{k-1}\right)^{'}e_k}{\sigma^2\nu+\left(e_k\right)^2}.\label{Apppsit}
\end{eqnarray}

Differentiating $\psi_k\left(e\right)$ with respect to $\hat{x}_1$ gives
\begin{eqnarray}
\frac{d\psi_k\left(e\right)}{d\hat{x}_1}&=&-\frac{d}{d\hat{x}_1}\left(\frac{(\nu+1)\left(H\Phi^{k-1}\right)^{'}e}{\sigma^2\nu+e^2}\right)\notag \\
&=&\left(H\Phi^{k-1}\right)^{'}H\Phi^{k-1}{(\nu+1)\left(\sigma^2\nu-e^2\right) \over \left(\sigma^2\nu+e^2\right)^2}. \label{Appdpsit}
\end{eqnarray}
Substituting \eqref{Apppsit} and \eqref{Appdpsit} into \eqref{IF} gives
\begin{eqnarray}
&&{\rm IF}(e)\notag \\
&=&\left(\textstyle{\sum_{k=1}^{T}\left(H\Phi^{k-1}\right)^{'}H\Phi^{k-1}\int_{-\infty}^{\infty}     }
          \frac{\left(\nu+1\right)\left(\sigma^2\nu-e^2\right)}
          {\left(\sigma^2\nu+e^2\right)^2}
              f\left(e\right)de\right)^{-1}\notag \\
              &&\times
        \left.\textstyle{\sum_{k=1}^{T}} \frac{\left(\nu+1\right)\left(H\Phi^{k-1}\right)^{'} e_k}{\sigma^2\nu+(e_k)^2}\right|_{\hat{x}_1=\bar x_1} \notag
\end{eqnarray}
which is \eqref{IFt}.

When $\nu=\infty$  we have
\begin{subequations}
\label{Appintnvinfty}
\begin{eqnarray}
&&\mathop{\rm lim}_{\nu\rightarrow \infty}\textstyle{\int_{-\infty}^\infty}{(\nu+1)\left(\sigma^2\nu-e^2\right) \over \left(\sigma^2\nu+e^2\right)^2}f(e)de=\frac{1}{\sigma^2}
\label{Appintnvinftya}
\end{eqnarray}
and
\begin{eqnarray}
&&\mathop{\rm lim}_{\nu\rightarrow \infty}{\left(\nu+1\right)\left(H\Phi^{k-1}\right)^{'}e_k\over \sigma^2\nu+e_k^2}=\frac{\left(H\Phi^{k-1}\right)^{'}e_k}{\sigma^2}. \label{Appintnvinftyb}
\end{eqnarray}
\end{subequations}
Substituting \eqref{Appintnvinftya} and \eqref{Appintnvinftyb} into \eqref{IFt} and setting $e_k=y_k-H\Phi^{k-1}\textcolor{black}{\bar x_1}-H\textcolor{black}{s}_k$ give
\begin{eqnarray}
{\rm IF}(e)&=&-\bar x_1+\left(\textstyle{\sum_{k=1}^T}\left(H\Phi^{k-1}\right)^{'}H\Phi^{k-1}\right)^{-1}\notag \\
&&\times\textstyle{\sum_{k=1}^{T}}\left(H\Phi^{k-1}\right)^{'}\left(y_k-H\textcolor{black}{s}_k\right)\notag
\end{eqnarray}
which is \eqref{IFgaussian}.
\section{Appendix: Proof of Theorem \ref{propMHETD}}
\label{APPpropMHETD}
Rewriting the state space model \eqref{armaxss} of ARMAX as
\begin{eqnarray}
x_{k+1}&=&\Phi x_k+\Gamma u_k+\Omega y_k \label{armaxxx} \\
y_k&=&Hx_k+e_k \notag
\end{eqnarray}
where $\Phi=\Phi_a-\Omega H$.
Then using \eqref{xIFxbar0} the current state estimate $x_{_{T}}$ is obtained by recursively applying the model \eqref{armaxxx}
\textcolor{black}{
\begin{eqnarray}
\label{xT1}
\hat x_{_{T}}&=&\Phi^{T-1}{\rm IF}(e)+\textstyle{\sum_{k=1}^{T-1}}\Phi^{T-k-1}\left(\Gamma u_k+\Omega y_k\right)\notag \\ &=&\textcolor{black}{s}_{_{T}}+\Phi^{T-1}{\rm {IF}}(e),
\end{eqnarray}}
where $s_{_T}$ is given by \eqref{xbar}.

Multiplying \eqref{IFarmaxMHE} by $\Phi^{T-1}$ {\color{black} and simplifying give
	\begin{eqnarray}
	&&\Phi^{T-1}{\rm IF}(e)\notag \\
	&=&\left(\textstyle{\sum_{\substack{i=1}}^{N}}\left(H\Phi^{i-N}\right)^{'}H\Phi^{i-N}\int_{-\infty}^\infty{(\nu+1)\left(\sigma^2\nu-e^2\right) \over \left(\sigma^2\nu+e^2\right)^2}f(e)de\right)^{-1}\notag \\
	&&\times\textstyle{\sum_{k=T-N+1}^{T}}{(\nu+1)\left(H\Phi^{k-T}\right)^{'}\left(y_k-H{s}_k\right) \over \sigma^2\nu+\left(y_k-H{s}_k\right)^2} \label{PhiIF}
	\end{eqnarray}
	
	}Rewriting \eqref{PhiIF} gives
\begin{eqnarray}
\Phi^{T-1}{\rm IF}(e)=\left(\textstyle{\sum_{i=1}^{N}}\left(H\Phi^{i-N}\right)'H\Phi^{i-N}\right)^{-1}\varPsi\mathcal W_{T,N},\notag \\
\label{IFMHE}
\end{eqnarray}
where
\begin{eqnarray}
\varPsi&=&\left[\begin{matrix}
\left(H\Phi^{1-N}\right)^{'}&\left(H\Phi^{2-N}\right)^{'}&\cdots&H^{'}
\end{matrix}\right], \notag\\
\mathcal W_{T,N}&=&\left[\begin{matrix}
w_{_{T-N+1}}&w_{_{T-N+2}}&\cdots&w_{_{T}}
\end{matrix}\right]^{'}, \notag\\
w_{_{T-N+i}}&=&\left(\textstyle{\int\limits_{-\infty}^{\infty}}\frac{\left(\nu+1\right)\left(\sigma^2\nu-e^2\right)}{\left(\sigma^2\nu+e^2\right)^2}f(e)de\right)^{-1}\notag \\
&&\times \frac{\left(\nu+1\right)\left(y_{_{T-N+i}}-H\textcolor{black}{s}_{_{T-N+i}}\right)}{\sigma^2\nu+\left(y_{_{T-N+i}}-H\textcolor{black}{s}_{_{T-N+i}}\right)^2},~i=1,2,\cdots,N. \notag
\end{eqnarray}
Equation \eqref{xhat} is obtained by substituting \eqref{IFMHE} into \eqref{xT1}.
\section{Appendix: Proof of Theorem \ref{theoremrecursiveMHET}}
\label{APPtheoremrecursiveMHET}
Define
\begin{eqnarray}
\label{deltax}
\Delta x_{_{T}}=\left(\textstyle{\sum_{i=1}^{N}}\left(H\Phi^{i-N}\right)'H\Phi^{i-N}\right)^{-1}\varPsi{\mathcal W}_{T,N}.
\end{eqnarray}
Thus the estimator \eqref{xhat} can be written as
\begin{eqnarray}
\hat x_{_{T}}=\textcolor{black}{s}_{_{T}}+\Delta x_{_{T}}.\notag
\end{eqnarray}
Notice that Equation \eqref{deltax} gives $\Delta x_{_T}$ from the minimization of the Moving Window Least-Squares loss function
\begin{eqnarray}
\notag
V=\frac{1}{2}\textstyle{\sum_{i=1}^{N}}\left(w_{_{T-N+i}}-H\Phi^{i-N}\Delta x_T)\right)^2.
\end{eqnarray}
The recursive solution of the above Moving Window Least-Squares \cite{ling1999receding} is
\begin{eqnarray}
\Delta x_{_{T+1}}&=&\left(\Phi-L_NH\Phi\right)\Delta x_{_{T}}+L_Nw_{_{T+1}}\notag \\
&&-\tilde{L}_N\left(w_{_{T-N+1}}-H\Phi^{-N+1}\Delta x_{_{T}}\right), \notag
\end{eqnarray}
where
\begin{eqnarray}
L_N&=&\Phi^{N-1}P_N\left(H\Phi^{N-1}\right)^{'}, \notag \\
\tilde{L}_N&=&\Phi^{N-1}P_N\left(H\Phi^{-1}\right)^{'}, \notag \\
P_N&=&\left(\textstyle{\sum_{i=1}^{N}}\left(H\Phi^{i-1}\right)^{'}H\Phi^{i-1}\right)^{-1}.  \notag
\end{eqnarray}
In the next time instant
\begin{eqnarray}
\hat{x}_{_{T+1}}
&=&\textcolor{black}{s}_{_{T+1}}+\Delta x_{_{T+1}} \notag \\
 &=&\Phi\textcolor{black}{s}_{_{T}}+\Gamma u_{_{T}}+\Omega y_{_{T}}+\left(\Phi-L_NH\Phi\right)\Delta x_{_{T}}\notag \\
 &&+L_Nw_{_{T+1}}-\tilde{L}_N\left(w_{_{T-N+1}}-H\Phi^{-N+1}\Delta x_{_{T}}\right) \notag \\
  &=&\Phi \hat{x}_{_{T}}+\Gamma u_{_{T}}+\Omega y_{_{T}}+L_N\left(w_{_{T+1}}-H\Phi \Delta x_{_{T}}\right)\notag \\
  &&-\tilde{L}_N\left(w_{_{T-N+1}}-H\Phi^{-N+1}\Delta x_{_{T}}\right). \label{xhatTT}
\end{eqnarray}
Expressing \eqref{xbar} in recursive form we have
 \begin{eqnarray}
 \label{xxbar}
\textcolor{black}{s}_{_{T+1}}=\Phi \textcolor{black}{s}_{_{T}}+\Gamma u_{_{T}}+\Omega y_{_{T}}
 \end{eqnarray}
 and
\begin{eqnarray}
\textcolor{black}{s}_{_{T-N+1}}=\Phi^{-N+1}\left(\textcolor{black}{s}_{_{T}}-\textstyle{\sum_{i=1}^{N-1}}\Phi^{i-1}\left(\Gamma u_{_{T-i}}+\Omega y_{_{T-i}}\right)\right). \notag \\
\label{xbarTN}
\end{eqnarray}
Define $z_k=w_k+H\textcolor{black}{s}_k.$
Substituting $w_{_{T+1}}=z_{_{T+1}}-H\textcolor{black}{s}_{_{T+1}}$ and $w_{_{T-N+1}}=z_{_{T-N+1}}-H\textcolor{black}{s}_{_{T-N+1}}$ into \eqref{xhatTT} gives
\begin{eqnarray}
\hat{x}_{_{T+1}}&=&\Phi \hat{x}_{_{T}}+\Gamma u_{_{T}}+\Omega y_{_{T}}\notag \\
&&+L_N\left(z_{_{T+1}}-H\textcolor{black}{s}_{_{T+1}}-H\Phi \Delta x_{_{T}}\right)\notag \\
&&-\tilde{L}_N\left(z_{_{T-N+1}}-H\textcolor{black}{s}_{_{T-N+1}}-H\Phi^{-N+1}\Delta x_{_{T}}\right). \notag \\
\label{xhatTT1}
\end{eqnarray}
Substituting \eqref{xxbar} and \eqref{xbarTN} into \eqref{xhatTT1}
gives
\begin{eqnarray}
&&\hat{x}_{_{T+1}}\notag \\
&=&\Phi \hat{x}_{_{T}}+\Gamma u_{_{T}}+\Omega y_{_{T}}+L_N\left(z_{_{T+1}}-H\left(\Phi\hat x_{_{T}}+\Gamma u_{_{T}}+\Omega y_{_{T}}\right)\right) \notag \\
&&-\tilde{L}_N\left(z_{_{T-N+1}}-H\Phi^{-N+1}\hat{x}_{_{T}}\right.\notag \\
&&\left.+H\Phi^{-N+1}\left(\varXi_uU_{T-1,N-1}+\varXi_yY_{T-1,N-1}\right)\right) \notag
\end{eqnarray}
which is Equation \eqref{Movingstrategy1}, and \eqref{Movingstrategy4} follows.

\section{Appendix: Proof of Theorem \ref{theorempropertyMHET}}
\label{APPtheorempropertyMHET}
	By substituting $y_k=Hx_k+e_k$ into \eqref{xbar} and assuming $x_1=0$ for simplicity, $s_{_T}$ can be rewritten as
	\begin{eqnarray}
	\textcolor{black}{s}_{_{T}}& = &\textstyle{\sum_{k=1}^{T-1}}\Phi_a^{T-k-1}\left(\Gamma u_{k}+\Omega e_{k}\right),~T\geqslant 2. \label{xbark}
	\end{eqnarray}
Substituting \eqref{xbark} into \eqref{xT1} gives
\begin{eqnarray}
\label{xhatapp}
\hat x_{_{T}}=\textstyle{\sum_{k=1}^{T-1}}\textcolor{black}{\Phi_a}^{T-k-1}\left(\Gamma u_k+\Omega e_k\right)+\Phi^{T-1}{\rm {IF}}(e).
\end{eqnarray}
Taking expectation of \eqref{xhatapp}, recognizing that $e_k$ is zero mean and $\textstyle{\int_{-\infty}^{\infty}}{\rm {IF}}(e)g(e)de=0$
from \eqref{IFarmaxMHE}, we get
\begin{eqnarray}
\notag
E\left(\hat x_{_{T}}\right)=\textstyle{\sum_{k=1}^{T-1}}\textcolor{black}{\Phi_a}^{T-k-1}\Gamma u_k.
\end{eqnarray}
 Therefore  the variance of $\hat x_{_{T}}$ is given by
 \begin{eqnarray}
 Var\left(\hat x_{_{T}}\right)&=&E\left(\left(\hat x_{_T}-E\left(\hat x_{_T}\right)\right)\left(\hat x_{_T}-E\left(\hat x_{_T}\right)\right)^{'}\right)\notag \\
 &=& \textstyle{\int_{-\infty}^{\infty}}\left(\textstyle{\sum_{k=1}^{T-1}}\textcolor{black}{\Phi_a}^{T-k-1}\Omega e_k+\Phi^{T-1}{\rm {IF}}(e)\right)\notag \\
 &&\times \left(\textstyle{\sum_{k=1}^{T-1}}\textcolor{black}{\Phi_a}^{T-k-1}\Omega e_k+\Phi^{T-1}{\rm {IF}}(e)\right)^{'}g(e)de. \notag \\
 \label{varyhatapp1}
 \end{eqnarray}
Substituting \eqref{IFarmaxMHE} into \eqref{varyhatapp1}, noticing that $e_k$ is i.i.d and simplifying,  give
 \begin{eqnarray}
Var\left(\hat x_{_{T}}\right)&=&\textstyle{\frac{\rho_1}{\rho_4^2}}M_N+2\textcolor{black}{\frac{\rho_2}{\rho_4}M_N}\textstyle{\sum_{i=1}^{N-1}}\left(H\Phi^{i-N}\right)^{'}\left(\textcolor{black}{\Phi_a}^{N-i-1}\Omega\right)^{'} \notag \\
&&+\textcolor{black}{\rho_3}\textstyle{\sum_{k=1}^{T-1}}\left(\textcolor{black}{\Phi_a}^{T-k-1}\Omega\right)\left(\textcolor{black}{\Phi_a}^{T-k-1}\Omega\right)^{'},
 \notag
 \end{eqnarray}
where
\begin{eqnarray}
\rho_1&=&\textstyle{\int_{-\infty}^{\infty}}\left(\frac{(\nu+1)e}{\sigma^2\nu + e^2}\right)^2g(e)de,~\rho_2=\textstyle{\int_{-\infty}^{\infty}}\frac{(\nu+1)e^2}{\sigma^2\nu + e^2}g(e)de, \notag \\
\rho_3&=&\textstyle{\int_{-\infty}^{\infty}}e^2g(e)de,~\rho_4=\textstyle{\int_{-\infty}^{\infty}}{(\nu+1)\left(\sigma^2\nu-e^2\right) \over (\sigma^2\nu+e^2)^2}g(e)de, \notag \\
M_N&=&\left(\textstyle{\sum_{i=1}^{N}}\left(H\Phi^{i-N}\right)^{'}H\Phi^{i-N}\right)^{-1}, \label{MN}
\end{eqnarray}
which is \eqref{varxhat}, and \eqref{varyhat} follows.
\section{Appendix: Proof of Theorem~\ref{corollaryMHEoutlier}}
\label{APPcorollaryMHEoutlier}
Recall \eqref{xbark} and $x_1=0$
\begin{eqnarray}
{s}_{_{T}}& = &\textstyle{\sum_{k=1}^{T-1}}{\Phi_a}^{T-k-1}\left(\Gamma u_k+\Omega e_k\right).\notag
\end{eqnarray}
Then
\begin{eqnarray}
E\left({s}_{_{T}}\right)=\left\{
\begin{aligned}
&\textstyle{\sum_{k=1}^{T-1}}{\Phi_a}^{T-k-1}\Gamma u_k~~~~~~~~~~~~{\rm if}~T\leq k_1, \\
&\textstyle{\sum_{k=1}^{T-1}}{\Phi_a}^{T-k-1}\Gamma u_k+{\Phi_a}^{T-k_1-1}\Omega e_{k_1} \notag\\
&~~~~~~~~~~~~~~~~~~~~~~~~~~~~~~~~~~~~~~~~{\rm if}~T>k_1,\notag
\end{aligned}
\right.
\end{eqnarray}

which is \eqref{meanxbaroutlier}. Let $k_2=k_1+N-1$.

Substituting \eqref{MN} into \eqref{PhiIF} gives
	\begin{eqnarray}
	\Phi^{T-1}{\rm IF}(e) &=&M_N\left(\int_{-\infty}^\infty{(\nu+1)\left(\sigma^2\nu-e^2\right) \over \left(\sigma^2\nu+e^2\right)^2}f(e)de\right)^{-1}\notag \\
	&&\times\textstyle{\sum_{k=T-N+1}^{T}}{(\nu+1)\left(H\Phi^{k-T}\right)^{'}\left(y_k-H{s}_k\right) \over \sigma^2\nu+\left(y_k-H{s}_k\right)^2} \label{PhiIF1}
	\end{eqnarray}
From the estimator \eqref{xT1} we have
\begin{eqnarray}
E\left(\hat x_{_{T}}\right)&=&E\left({s}_{_{T}}\right)+\textstyle{\int_{-\infty}^{\infty}}\Phi^{T-1}{\rm IF}(e)g(e)de. \label{meanxhatapp}
\end{eqnarray}
Using \eqref{meanxhatapp} and \eqref{yhat} gives
\begin{eqnarray}
E\left(\hat y_{_{T}}\right)&=&HE\left({s}_{_{T}}\right)+H\textstyle{\int_{-\infty}^{\infty}}\Phi^{T-1}{\rm {IF}}(e)g_k(e)de. \label{meanyhatapp2}
\end{eqnarray}

Substituting \eqref{PhiIF1} and \eqref{gkpdf} into the second term of \eqref{meanyhatapp2} and simplifying give

	\begin{eqnarray}
	&&H\textstyle{\int_{-\infty}^{\infty}}\Phi^{T-1}{\rm {IF}}(e)g_k(e)de\notag \\
	&=&\left\{
	\begin{aligned}
	&HM_N\left(\textstyle{\int_{-\infty}^{\infty}}\frac{\left(\nu+1\right)\left(\sigma^2\nu-e^2\right)}{\left(\sigma^2\nu+e^2\right)^2}f(e)de\right)^{-1}\notag \\
	&~~~\times\textstyle{\int_{-\infty}^{\infty}}{\frac{\left(\nu+1\right)\left(H\Phi^{k_1-T}\right)^{'}e}{\sigma^2\nu+e^2}\Delta_{e_{k_1}}de} \notag\\
	&~~~{+}HM_N\left(\textstyle{\int_{-\infty}^{\infty}}\frac{\left(\nu+1\right)\left(\sigma^2\nu-e^2\right)}{\left(\sigma^2\nu+e^2\right)^2}f(e)de\right)^{-1}\notag \\
	&~~~\textstyle{\int_{-\infty}^{\infty}\sum_{\substack{i=1\\i\neq {k_1-T+N}} }^{N}}\frac{\left(\nu+1\right)\left(H\Phi^{i-{N}}\right)^{'}e}{\sigma^2\nu+e^2}g(e)de\notag \\
	&~~~~~~~~~~~~~~~~~~~~~~~~~~~~~~~~~~~~~~~~~~~~~~{\rm if}~k_1\leq T \leq k_2,  \\
	& \\
	&0~~~~~~~~~~~~~~~~~~~~~~~~~~~~~~~~~~~~~~{\rm if}~T<k_1~{\rm or}~T>k_2.
	\end{aligned}\right. \notag \\
	\label{intIF2}
	\end{eqnarray}
With
\begin{eqnarray}
\notag
\textstyle{\int_{-\infty}^{\infty}\sum_{\substack{i=1\\i\neq k_1-T+N} }^{N}}\frac{\left(\nu+1\right)\left(H\Phi^{{i}-N}\right)^{'}e}{\sigma^2\nu+e^2}g_k(e)de=0,
\end{eqnarray}
equation \eqref{intIF2} becomes
\begin{eqnarray}
&&H\textstyle{\int_{-\infty}^{\infty}}\Phi^{T-1}{\rm {IF}}(e)g(e)de\notag \\
&=&\left\{
\begin{aligned}
& HM_N\left(\textstyle{\int_{-\infty}^{\infty}}\frac{\left(\nu+1\right)\left(\sigma^2\nu-e^2\right)}{\left(\sigma^2\nu+e^2\right)^2}f(e)de\right)^{-1}\notag \\
&~~~\textstyle{\int_{-\infty}^{\infty}}{\frac{\left(\nu+1\right)\left(H\Phi^{k_1-T}\right)^{'}e}{\sigma^2\nu+e^2}\Delta_{e_{k_1}}de} ~~~{\rm if}~k_1\leq T \leq k_2,  \\
& \\
&0~~~~~~~~~~~~~~~~~~~~~~~~~~~~~~~~~~~~~~{\rm if}~T<k_1~{\rm or}~T>k_2,
\end{aligned}\right.  \notag \\
&=&\left\{
\begin{aligned}
&\textstyle{\frac{1}{\rho_4}}HM_N\frac{\left(H\Phi^{k_1-T}\right)^{'}(\nu+1)e_{k_1}}{\sigma^2\nu+e_{k_1}^2}~~~~~~~~~{\rm if}~k_1\leq T \leq k_2,  \\
&0~~~~~~~~~~~~~~~~~~~~~~~~~~~~~~~~~~~~~~{\rm if}~T<k_1~{\rm or}~T>k_2.
\end{aligned}\right. \notag \\
\label{intIF3}
\end{eqnarray}

Equation \eqref{meanyhatoutlier} is obtained by substituting \eqref{intIF3} \replaced{\textcolor{black}{into}}{} \eqref{meanyhatapp2}.

\section{Appendix: Numerical Illustration Example}
\label{APPExample}

  For example, for $N=6$ and $k=9$, substituting the parameters for Example 1 into \eqref{varyhat} gives

  \begin{eqnarray}
  M_N&=&\left(\sum_{i=1}^{6}\left(\left[\begin{matrix}
  1&0&0&&0&0
  \end{matrix}\right]\left[\begin{matrix}
  -4.0000&1&0&0&0\\
  -6.4000&0&1&0&0\\
  -5.1200&0&0&1&0\\
  -2.0480 &0&0&0&1\\
  -0.3278&0&0&0&0
  \end{matrix}\right]^{i-6}\right)^{'}\right.\notag \\
  &&\left.\times\left[\begin{matrix}
  1&0&0&&0&0
  \end{matrix}\right]\left[\begin{matrix}
  -4.0000&1&0&0&0\\
  -6.4000&0&1&0&0\\
  -5.1200&0&0&1&0\\
  -2.0480 &0&0&0&1\\
  -0.3277&0&0&0&0
  \end{matrix}\right]^{i-6}\right)^{-1} \notag \\
  &=&\left[\begin{matrix}
  0.9887&0.7858&0.3986&0.1163&0.0148\\
  0.7858&16.8432&16.2135&6.6997&1.0672\\
  0.3986&16.2135&16.4601&7.0558&1.1553\\
  0.1163&6.6997&7.0558&3.1051&0.5187\\
  0.0148&1.0672&1.1553&0.5187&0.0880
  \end{matrix}\right] \notag
  \end{eqnarray}
  and
  \begin{eqnarray}
  Var\left(\hat y_{_T}\right)=0.9887\frac{\rho_1}{\rho_4^2}+0.0064\frac{\rho_2}{\rho_4}+0.4481\rho_3=0.7055, \notag
  \end{eqnarray}
  $\rho_1=2.6667,~\rho_2=1,~\rho_3=0.7414,~\rho_4=2.6667$.
  Substituting the parameters for Example 1 into \eqref{varyhatgaussian} gives
  \begin{eqnarray}
  Var\left(\hat y_{_T}\right)=\rho_3\left(0.9887+0.0064+0.4481\right)=1.0705. \notag
  \end{eqnarray}

\end{appendices}

\fi

\end{document}